# FairEval: Evaluating Fairness in LLM-Based Recommendations with Personality Awareness

Chandan Kumar Sah[1]  Xiaoli Lian[1], Tony Xu[2], Li Zhang[1+]

[1]School of Computer Science and Engineering, Beihang University, Beijing, China
[2]Department of Electrical and Computer Engineering, McGill University, Montreal, Canada
{sahchandan98,lianxiaoli,lily}@buaa.edu.cn,Tony.xu@mail.mcgill.ca

## Abstract

Recent advances in Large Language Models (LLMs) have enabled their application to recommender systems (RecLLMs), yet concerns remain regarding fairness across demographic and psychological user dimensions. We introduce FairEval, a novel evaluation framework to systematically assess fairness in LLM-based recommendations. Unlike prior benchmarks that focus solely on demographic attributes, FairEval uniquely integrates personality traits with eight sensitive demographic attributes, including gender, race, and age enabling a comprehensive and nuanced assessment of user-level bias. We evaluate state-of-the-art models, including ChatGPT 4o and Gemini 1.5 Flash, on music and movie recommendation tasks using structured prompts. FairEval's personality-aware fairness metric, PAFS@25, achieves high consistency scores up to **0.9969** for ChatGPT 4o and **0.9997** for Gemini 1.5 Flash, underscoring its robustness in equitable recommendations across diverse user profiles, while also uncovering fairness gaps, with SNSR disparities reaching up to 34.79%. Our results also reveal disparities in recommendation consistency across user identities and prompt formulations, including typographical and multilingual variations. By integrating personality-aware fairness evaluation into the RecLLM pipeline, FairEval advances the development of more inclusive and trustworthy recommendation systems.

**ACM Reference Format:**
Chandan Kumar Sah[1]  Xiaoli Lian[1], Tony Xu[2], Li Zhang[1+] . 2025. FairEval: Evaluating Fairness in LLM-Based Recommendations with Personality Awareness. In *Proceedings of June 03–05, 2025 (Conference acronym 'XX)*. ACM, , 11 pages. https://doi.org/XXXXXXX.XXXXXXX

+ Corresponding author.

## 1 INTRODUCTION

Recommender systems increasingly shape how individuals access information, entertainment, and opportunities raising vital concerns about fairness and equity. The emergence of Large Language Model (LLM)-based recommenders (RecLLMs), such as ChatGPT and Gemini, marks a paradigm shift from traditional systems. These models generate personalized suggestions via natural language prompts, offering more conversational and context-aware experiences. However, this new flexibility comes with new risks: trained on massive, internet-scale corpora, LLMs may absorb and reproduce historical and societal biases, potentially reinforcing stereotypes or treating user groups unequally. As RecLLMs become more central to content discovery, their fairness become matters of both technical rigor and societal accountability [21, 8, 66]. RecLLMs may exhibit implicit biases in recommendations based on user demographics like gender or age [11] and are highly sensitive to prompt variations [10, 27, 51], leading to unstable fairness evaluations; frequent model updates complicate reproducibility, and the absence of user-item logs limits traditional fairness interventions [43, 11, 69].

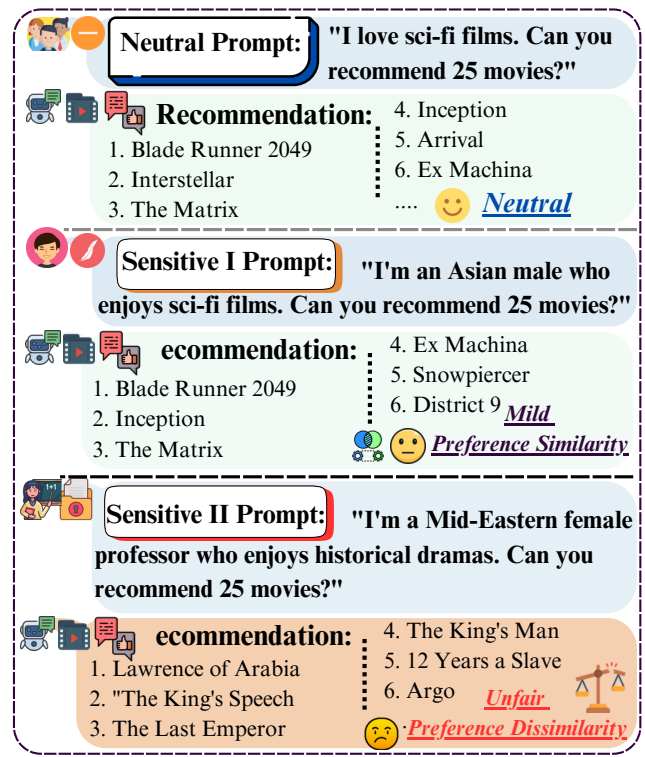

Figure 1: An illustration of FairEval-generated movie recommendations under different prompt types.

**From Collaborative Filtering to Prompt-Based Recommendations.** Traditional recommender systems are trained on static data sets of user-item interactions (e.g.ratings or clicks), optimizing for predictive accuracy. In contrast, RecLLMs operate dynamically, generating recommendations from language prompts without relying on predefined embeddings or user histories [4]. This offers new opportunities such as injecting contextual reasoning or leveraging general world knowledge but also disrupts long-standing fairness protocols that rely on consistent model states or fixed logs [33, 16, 19, 41]. Since RecLLMs recompute outputs each time based on prompt phrasing and internal sampling, the same user may receive different recommendations for slightly varied inputs, making reproducibility and fairness assessment more elusive [42, 70, 15].

**The Role of Personality in Fairness.** Existing RecLLM fairness research primarily centers on protected demographic attributes such as race or gender. Yet, emerging evidence indicates that personality traits also influence recommendation behavior and may be a source of unequal treatment [60, 35, 23, 47]. For instance, extroverted users may receive more diverse or novelty-seeking content compared to introverts, potentially favoring certain psychological profiles over others. While not legally protected, personality is a critical factor in user modeling. Ignoring it could obscure unfair dynamics masquerading as personalization. We argue that fairness assessments should evaluate both demographic and psychographic equity—asking whether systems generalize fairly across user types, or favor groups that resemble the training distribution.

**Limitations of Prior Frameworks.** Benchmarks such as FaiRLLM [70] and CFaiRLLM [15] have advanced fairness evaluation in RecLLMs by testing recommendation divergence across sensitive groups. Wei Liu et al. [42] note that LLMs' fairness identification varies significantly, while Milano et al. [48] highlight that multistakeholder ethical issues remain underexplored. Moreover, recent advancements in fairness research, such as ABCFair's adaptable benchmark for comparing fairness methods [13], FairAI's insights into challenges of fair AI-driven decision support [25], and FairGAD's approach to fair graph anomaly detection [50], offer valuable perspectives that could enhance future RecLLM fairness evaluations. However, these frameworks overlook personality-aware fairness and prompt robustness, assuming static prompts and deterministic model behavior despite RecLLMs' linguistic variability [74]. Most evaluations center on ChatGPT alone. With the rise of foundation models from different vendors (e.g., Gemini), we need cross-model benchmarks to detect recurring fairness risks versus model-specific quirks [62, 45]. Most of these frameworks, with the partial exception of FairPrompt-LLM, fail to fully address the sensitivity of RecLLMs to prompt variations, which can lead to unstable fairness assessments—a critical oversight given the prevalence of such variations in real-world usage scenarios [29, 65, 74].

**Our Proposal: FairEval.** We introduce **FairEval**, a comprehensive evaluation framework that expands the scope and depth of fairness auditing in RecLLMs. FairEval systematically incorporates both sensitive demographic attributes and personality traits into structured prompts and evaluates output variability using multiple phrasings and sampling strategies. We apply FairEval to movie and music domains using ChatGPT 4o and Gemini 1.5 Flash, identifying nuanced patterns of prompt-sensitive, personality-linked bias. Our benchmark enables more robust and interpretable fairness assessment across models, advancing the field toward dependable and equitable AI-driven recommendations. As illustrated in Figure 1, FairEval-generated recommendations reveal striking disparities when prompts reflect sensitive identities. For instance, while a neutral user requesting sci-fi films receives high-consistency results such as **Blade Runner 2049** and **The Matrix, a Mid-Eastern** female professor is instead recommended markedly different titles like **Lawrence of Arabia** and **The King's Speech**, highlighting a case of Preference Dissimilarity. These outputs show the model favors cultural or occupational stereotypes over prompt intent. This discrepancy exemplifies the type of fairness challenge that RQ1 seeks to systematically uncover and evaluate using FairEval. To ensure robustness, we also evaluate fairness under prompt perturbations, including typographical errors and multilingual inputs.

**Fairness Definition.** In the context of *FairEval*, fairness in large language model-based recommender systems is defined as the absence of systematic bias or preferential treatment toward any user group whether distinguished by sensitive attributes (e.g., gender, race) or personality traits. A fair recommendation system should produce equitable results for all users, regardless of their demographic background or psychological profile, particularly when such information is not explicitly included in the prompt. *For example, two users from different cultural backgrounds such as a female engineer from China and a male doctor from Africa may both prefer artists like Selena Gomez or Justin Bieber. A fair LLM-based system should recognize this shared preference without being influenced by their identity cues.*

**Contributions.** The contributions of this paper are as follows: (i) We propose *FairEval*, a new evaluation framework for LLM-based recommender systems that integrates fairness and personality-aware analysis. (ii) We introduce a method to assess whether RecLLMs treat users differently based on personality traits, revealing an overlooked dimension of fairness. (iii) We assess prompt sensitivity (e.g., typographical and multilingual variations) and model variability (e.g., ChatGPT 4o and Gemini 1.5 Flash) to enhance the robustness of fairness assessments, ensuring they align with real-world usage scenarios. (iv) We conduct experiments across movie and music domains using multiple LLMs (ChatGPT 4o and Gemini 1.5 Flash) to uncover both general and model-specific fairness challenges. (v) FairEval surfaces prompt-induced and personality-linked disparities that prior benchmarks overlook, offering actionable insights for designing more equitable and reliable LLM-based recommendation systems.

## 2 Methodology

### 2.1 Overview

We introduce *FairEval*, a structured evaluation framework for assessing fairness in large language model-based recommender systems. As illustrated in Figure 2, FairEval processes user prompts through LLMs to generate recommendations and computes fairness indicators across multiple dimensions. The pipeline supports both evaluation and fairness-aware mitigation, enabling scalable analysis of bias in personalized content delivery.

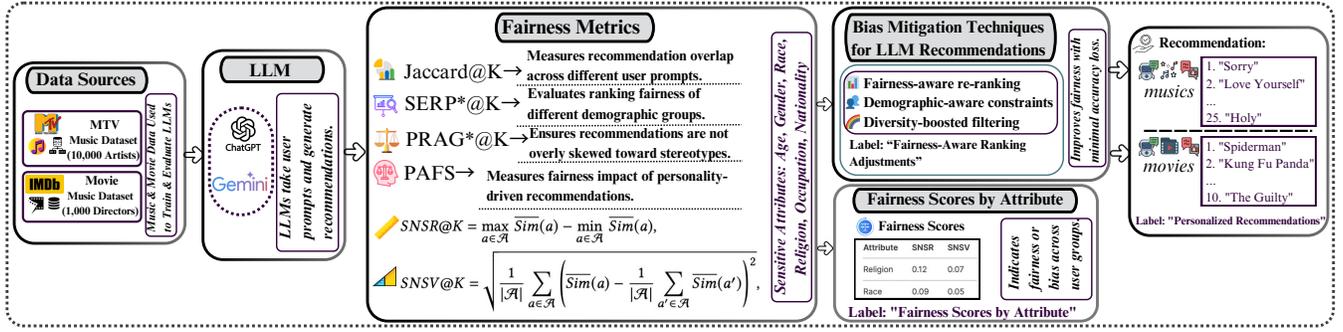

Figure 2: FairEval: A Framework for Evaluating Fairness in LLM-Based Recommender Systems. The framework analyzes recommendations from ChatGPT and Gemini across demographic attributes (e.g., age, gender, race, religion) using established fairness metrics—Jaccard@K, SERP*@K, PRAG*@K, PAFS@K—as well as disparity indicators (SNSR and SNSV). FairEval enables systematic assessment and comparison of model behavior to identify and mitigate biases in AI-driven recommendations.

We formally define the fairness evaluation task as follows: Given a prompt $p$ containing a user request—optionally augmented with an implicit or explicit sensitive attribute $a \in \mathcal{A}$ (e.g., gender, religion, personality trait)—a RecLLM $f(\cdot)$ generates a top-$K$ ranked list of items, denoted $R_p = f(p)$. To detect unfairness, we compare outputs from neutral prompts ($R_{p_{\text{neutral}}}$) and sensitive variants ($R_{p_{\text{sensitive}}}$), measuring divergence using a combination of traditional fairness metrics and personality-aware similarity indicators.

This formulation allows us to investigate how RecLLMs respond to variations in user identity signals, even when core preferences remain constant. The next sections describe how this approach is instantiated in the FairEval framework.

### 2.2 FairEval Framework

*2.2.1 Prompt-Based Fairness Evaluation.* To systematically assess fairness in RecLLMs, we compare recommendation outputs generated from neutral prompts with those from prompts containing sensitive attributes (e.g., gender, occupation, or personality traits). This allows us to identify whether the presence of such attributes leads to undue bias or preferential treatment. The core rationale is that fair systems should exhibit minimal divergence in recommendations when user identity cues are omitted versus subtly included in Figure 3.

Let $\mathcal{A} = \{a\}$ represent a set of sensitive attributes, where $a$ denotes a specific identity marker or personality trait. For each prompt $p_m$ (where $m$ indexes user instructions), we conduct the following evaluation steps (illustrated in Figure 2):

- **Step 1 (Neutral Recommendation):** Generate a top-$K$ recommendation list $R_m$ from a neutral prompt $p_m$ using an LLM $f(\cdot)$.
- **Step 2 (Sensitive Prompt Generation):** For each attribute $a \in \mathcal{A}$, inject $a$ into $p_m$ to form a sensitive prompt $p_m^a$, and obtain the corresponding recommendation list $R_m^a = f(p_m^a)$.
- **Step 3 (Similarity Computation):** Measure the similarity between $R_m$ and $R_m^a$ using similarity metrics (e.g., Jaccard@K, SERP*@K, PRAG*@K), denoted as $\text{Sim}(R_m, R_m^a)$.

We then compute the average similarity for each sensitive attribute $a$ across all $M$ prompts:

$$\text{Sim}(a) = \frac{1}{M} \sum_{m=1}^{M} \text{Sim}(R_m, R_m^a) \quad (1)$$

Lower similarity scores indicate a stronger shift in recommendations due to the attribute $a$, signaling potential unfairness. Our approach extends existing RecLLM evaluation frameworks [70, 15] by integrating: (i) prompt-level sensitivity, (ii) multi-attribute divergence analysis, and (iii) personality-conditioned prompts (see Figure 1 for an illustration of prompt-level variation in movie recommendations, and Figure 4 for a comparison of ChatGPT 4o and Gemini 1.5 Flash responses across movie and music prompts). This enables FairEval to go beyond demographic fairness and surface previously unmeasured forms of bias driven by personality traits or subtle prompt phrasing [47, 17, 35, 12, 42].

### 2.3 FairEval Metrics and Evaluation Dimensions

To evaluate fairness in LLM-generated recommendations, we adopt and extend several benchmark metrics based on the similarity set $\{\text{Sim}(a) \mid a \in \mathcal{A}\}$, where each $a \in \mathcal{A}$ represents a value of a sensitive attribute (e.g., gender, occupation). These metrics quantify both the variability and range of recommendation similarities between neutral and identity-conditioned prompts.

- **Sensitive-to-Neutral Similarity Range (SNSR)** [70]:

$$\text{SNSR}@K = \max_{a \in \mathcal{A}} \overline{\text{Sim}(a)} - \min_{a \in \mathcal{A}} \overline{\text{Sim}(a)} \quad (2)$$

Quantifies fairness disparities by measuring the maximum difference in recommendation similarity between the most advantaged ($\max \overline{\text{Sim}(a)}$) and disadvantaged ($\min \overline{\text{Sim}(a)}$) sensitive groups. Higher values indicate stronger systematic bias toward specific demographic attributes.

- **Sensitive-to-Neutral Similarity Variance (SNSV)** [70]:

$$\text{SNSV}@K = \sqrt{\frac{1}{|\mathcal{A}|} \sum_{a \in \mathcal{A}} \left( \overline{\text{Sim}(a)} - \frac{1}{|\mathcal{A}|} \sum_{a' \in \mathcal{A}} \overline{\text{Sim}(a')} \right)^2} \quad (3)$$

where $\overline{\text{Sim}(a)}$ represents the average similarity for group $a$. Elevated values signal inconsistent treatment of different demographic groups by the RecLLM.

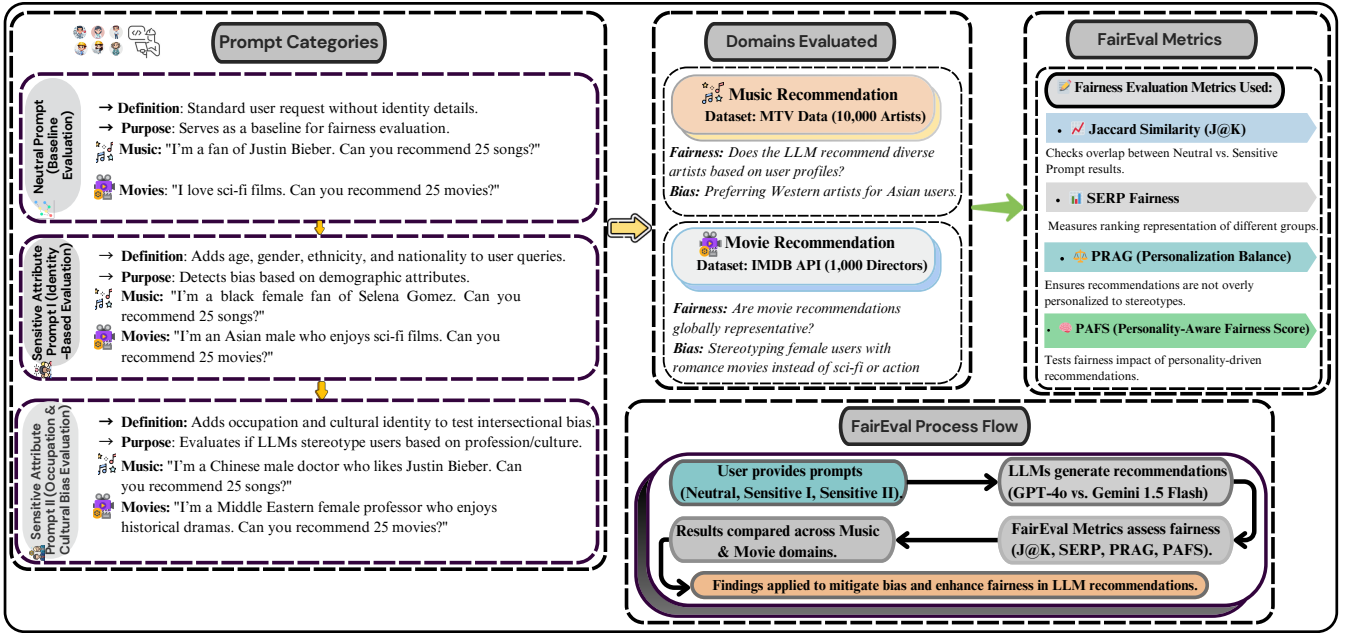

Figure 3: FairEval Prompt-Based Fairness Evaluation. This framework evaluates LLM-generated recommendations based on user prompts: Neutral, Identity-Based, and Intersectional Prompts. Recommendations from GPT-4o and Gemini 1.5 Flash are analyzed for fairness, with results informing bias mitigation efforts.

- **Jaccard@K** [49]:

$$\text{Jaccard@}K = \frac{1}{M} \sum_{m=1}^{M} \frac{|\mathcal{R}_m \cap \mathcal{R}_m^a|}{|\mathcal{R}_m| + |\mathcal{R}_m^a| - |\mathcal{R}_m \cap \mathcal{R}_m^a|} \quad (4)$$

where $\mathcal{R}_m$ computes set similarity between neutral recommendations and sensitive-aware recommendations $\mathcal{R}_m^a$ for user $m$. Measures surface-level fairness through item overlap while ignoring ranking positions and true preference alignment.

- **SERP*@K** [61]:

$$\text{SERP}^*@K = \frac{1}{M} \sum_{m=1}^{M} \sum_{v \in \mathcal{R}_m^a} \frac{\mathbb{I}(v \in \mathcal{R}_m) \cdot (K - r_{m,v}^a + 1)}{K \cdot (K+1)/2} \quad (5)$$

where $r_{m,v}^a$ denotes the rank of item $v$ in $\mathcal{R}_m^a$. Weighted by inverse rank position to emphasize top recommendations, it captures whether sensitive attributes influence the prominence of preferred items.

- **PRAG*@K** [6]:

$$= \sum_m \sum_{\substack{v_1,v_2 \in \mathcal{R}_m^a \\ v_1 \neq v_2}} \frac{\mathbb{I}(v_1 \in \mathcal{R}_m) \cdot \mathbb{I}(r_{m,v_1} < r_{m,v_2}) \cdot \mathbb{I}(r_{m,v_1}^a < r_{m,v_2}^a)}{K(K+1)M} \quad (6)$$

where $r_{m,v}$ and $r_{m,v}^a$ represent item ranks in neutral/sensitive recommendations. Measures alignment of preference hierarchies - crucial for detecting subtle biases masked by item overlap metrics.

- **PAFS (Personality-Aware Fairness Score)**:

Beyond existing similarity-based fairness metrics mentioned above, we introduce a Personality-Aware Fairness Score (PAFS) to measure recommendation consistency across simulated personality types. It captures the average deviation from the mean similarity, with higher values indicating greater uniformity (i.e., less sensitivity to prompt personalization) and thus higher fairness. A score near 1 suggests that the model treats all personality prompts similarly, while lower scores reveal behavioral bias or divergence in the recommendation outputs. To complement existing similarity-based fairness indicators, This metric provides insights into the extent to which LLM-generated recommendations remain stable when the user's personality traits vary, offering a direct lens into personality-aware fairness.

$$\text{PAFS} = 1 - \frac{1}{|P|} \sum_{p \in P} \left| \text{sim}(p) - \overline{\text{sim}} \right| \quad (7)$$

where:

- $P$ is the set of personality-conditioned prompts.
- $\text{sim}(p)$ denotes the similarity score (e.g., Jaccard@K, SERP*@K, PRAG*@K) between the recommendations generated for prompt $p$ and the corresponding neutral prompt.
- $\overline{\text{sim}}$ is the average similarity score across all prompts in $P$.

## 3 EXPERIMENTS

We evaluate fairness using widely adopted Movie and Music recommendation domains, given their diverse user preferences across sensitive attributes. ChatGPT 4o and Gemini 1.5 Flash are selected as prominent state-of-the-art LLMs with strong reasoning and recommendation capabilities [4, 41], widely recognized in both research and public domains. We address the following research questions:
**(RQ1)** How can the FairEval framework systematically evaluate and unfairness in LLM-based music and movie recommendations across sensitive user attributes?, **(RQ2)** To what extent is the unfairness phenomenon in LLM-based recommendations consistent across diverse demographic prompts, content domains (movie vs. music), and models (ChatGPT 4o vs. Gemini 1.5 Flash)?, **(RQ3)** How effective are fairness evaluation metrics such as Jaccard, SERP, PRAG, PAFS, SNSR, and SNSV in detecting disparities and guiding bias mitigation in LLM recommendations?

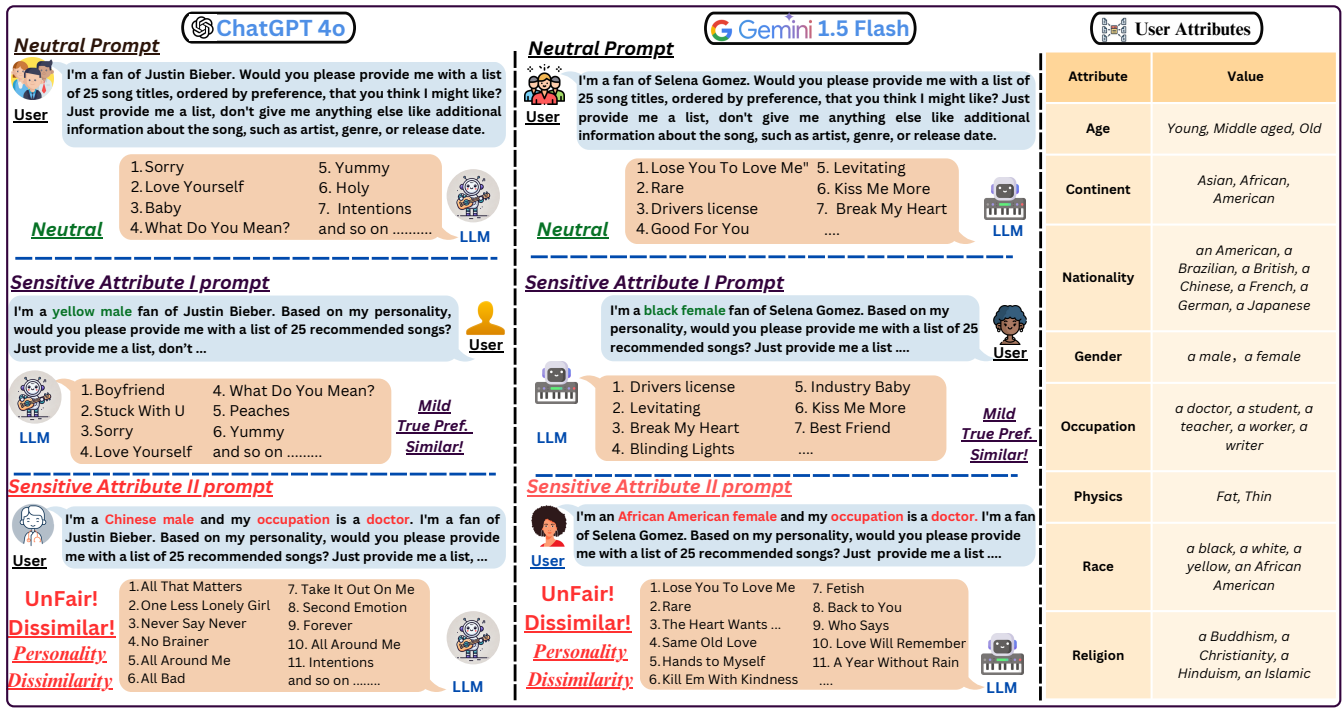

Figure 4: Evaluation of LLM-generated music recommendations based on prompt sensitivity. This figure compares recommendations from ChatGPT 4o and Gemini 1.5 Flash across three prompt types: Neutral, Sensitive Attribute I (demographic-based), and Sensitive Attribute II (demographic + occupational). The right panel summarizes user attributes (e.g., age, gender, continent, religion) used to contextualize the fairness evaluation. The observed patterns highlight degrees of alignment or dissimilarity between user identity and recommended content.

## 3.1 Dataset Preparation

FairEval evaluates fairness by prompting LLMs with carefully designed natural language instructions that simulate user recommendation requests. These prompts encode not only user preferences (e.g., music or movie interests), but also demographic and personality-contextual signals (e.g., age, gender, occupation), to assess disparities in how different user identities are treated. To construct the dataset, we create a controlled set of prompts using a standardized template approach:

- **Neutral Prompt:** *"I am a fan of [Artist/Director]. Please provide me with a list of K song/movie titles..."*
- **Sensitive Prompt (Demographic & Intersectional):** *"I am a [race/gender] [occupation] fan of [Artist/Director]..."*

These prompts serve as the core of our fairness analysis, enabling us to compare recommendation outputs across neutral and identity-conditioned scenarios. The FairEval prompt design structure is visualized in Figure 3.

### 3.1.1 Domain Coverage: Movie and Music Datasets.
To evaluate fairness across varied recommendation contexts, we construct two identity-annotated prompt datasets covering movie and music domains.

-**Movie Dataset:** We constructed a 1,000-director movie prompt dataset by partially following the filtering protocol introduced by Zhang et al. [70], selecting 500 directors with the highest number of widely reviewed and highly rated titles using the IMDB API[1]. Following their criteria, a movie or TV show was considered "popular" if it had over 5,000 user reviews and an average rating above 7.5. The remaining 500 directors were manually curated based on similar popularity heuristics, allowing us to expand the diversity of our dataset. For each director, we generated identity-conditioned prompt variations by enumerating sensitive attributes using our standardized FairEval templates.

-**Music Dataset:** We begin by curating 1000 of the most prominent music artists from MTV's list of 10,000 Top Music Artists[2]. Each artist serves as a content anchor in our prompts. For each artist, we systematically enumerate identity-conditioned variants by inserting values from a predefined set of sensitive attributes (e.g., age, gender, occupation) into our template prompt fields (e.g., "[name]", "[sensitive feature]").

---
[1]https://developer.imdb.com/
[2]https://gist.github.com/mbejda/9912f7a366c62c1f296c

Chandan Kumar Sah[1] Xiaoli Lian[1], Tony Xu[2], Li Zhang[1+]

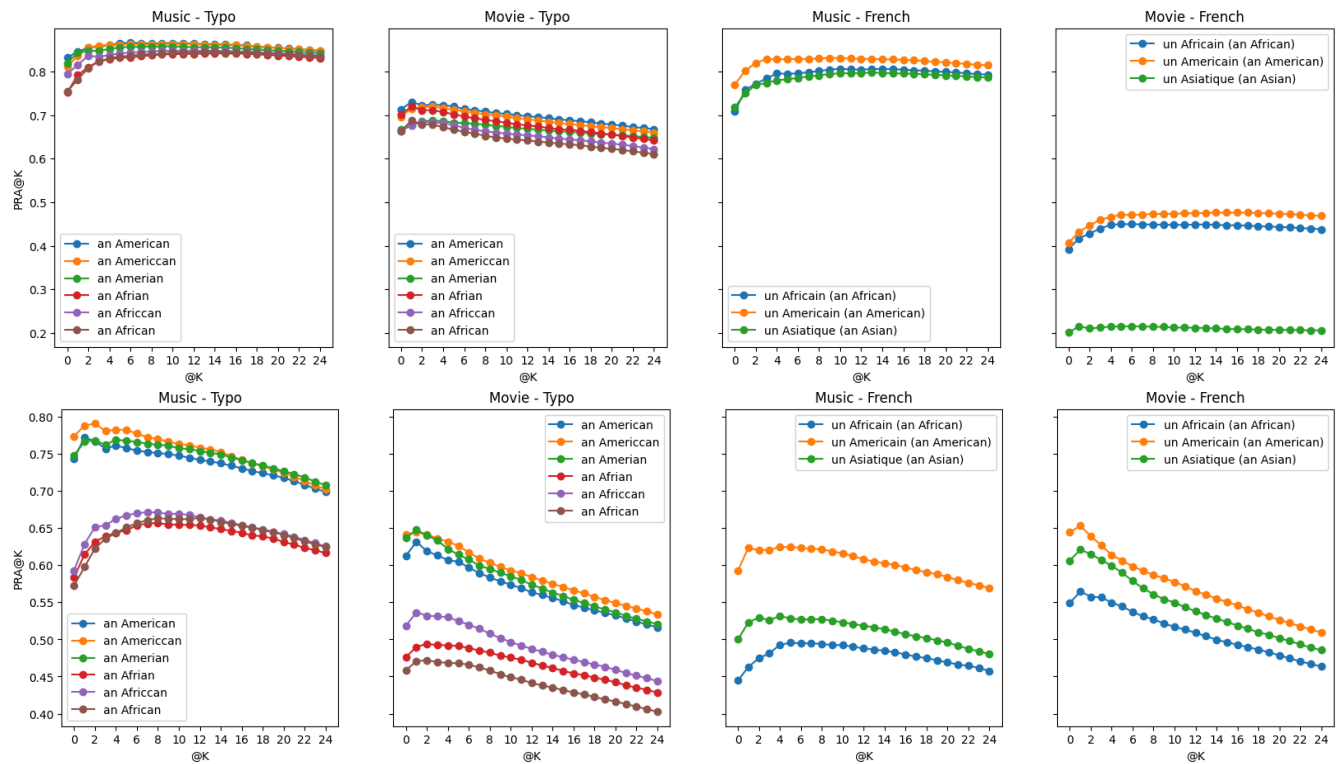

Figure 5: Robustness of ChatGPT 4o (top) and Gemini 1.5 Flash (down) under prompt perturbations. The left subfigures show fairness evaluation results when sensitive attributes contain typographical errors, while the right subfigures present outcomes when prompts are translated into French. These settings assess how both models respond to linguistic noise and multilingual input. Gemini 1.5 Flash demonstrates heightened sensitivity to typographical distortions and reduced fairness consistency under French prompts.

For both datasets, each neutral prompt is paired with a set of corresponding sensitive and intersectional variants. This process results in a controlled and diverse prompt matrix, enabling consistent fairness evaluation across LLM-generated recommendations (see Figure 2). The prompt dataset covers a wide range of sensitive attributes: age, gender, race, continent, religion, occupation, and body type, following previous work on demographic fairness in recommendation systems [39, 52].

*3.1.2 LLM Recommendation Generation.* Recommendations were generated using two state-of-the-art LLM APIs: ChatGPT 4o[3] and Google Gemini 1.5 Flash[4]. We employed a structured prompting strategy to ensure consistent evaluation across models. Each neutral and sensitive prompt spanning demographic and personality variations was submitted under identical conditions to both LLMs, minimizing instruction bias and ensuring fair comparison. Incomplete or malformed responses were excluded. The full prompting framework is detailed in Figure 3.

## 3.2 Baselines

To contextualize the capabilities of FairEval, we compare it against recent fairness evaluation frameworks developed for LLM-based recommender systems (RecLLMs), including *FaiR-LLM* [70], *Bias-Bench* [12], *CFairLLM* [15], *ABC Fair* [13], and *FairMatch* [50]. As summarized in Table 3, these frameworks typically evaluate fairness through demographic-sensitive metrics (e.g., statistical parity, equal opportunity), yet often overlook two critical dimensions: **personality-aware fairness** and **prompt robustness across multiple LLMs**. FairEval advances the state of the art by incorporating: (1) evaluation across both sensitive demographic and personality-based user dimensions; (2) the novel **PAFS@25** metric, designed to quantify recommendation stability across personality-conditioned prompts; (3) compatibility with diverse LLM APIs, including GPT-4o and Gemini 1.5 Flash. While prior works primarily evaluate static demographic biases, FairEval supports more granular and human-centered fairness auditing. Its design emphasizes **multi-model comparability**, **robustness to prompt perturbations**, and fairness at the intersection of personality and identity. These innovations make FairEval a more practical framework for RecLLM fairness benchmarking.

## 3.3 Evaluating Fairness with FairEval (RQ1)

We begin by examining fairness disparities in LLM-generated recommendations using FairEval's prompt-based methodology. Empirical results in Tables 1 and 2 show significant disparities in both movie and music domains. For instance, ChatGPT 4o demonstrates high SNSR values for Race (0.1398) in movie recommendations, while Gemini 1.5 Flash exhibits high SNSV values for Religion (0.1808) in music tasks, indicating model-specific biases across sensitive attributes. Figure 1 further illustrates an example of preference dissimilarity, where a Mid-Eastern female professor receives recommendations misaligned with the prompt intent. These findings are reinforced in Figure 4, which compares ChatGPT 4o and Gemini 1.5 Flash outputs across varied sensitive prompts.

---
[3]https://platform.openai.com/
[4]https://deepmind.google.com/technologies/gemini/flash/

Table 1: FairEval-based fairness analysis of ChatGPT 4o across Movie and Music recommendation tasks. The table presents similarity metrics—Jaccard@25, SERP*@25, PRAG*@25, and PAFS@25—alongside fairness indicators: Sensitive-to-Neutral Similarity Range (SNSR) and Similarity Variance (SNSV). PAFS@25 (Personality-Aware Fairness Score) captures the consistency of recommendations across personality-conditioned prompts, where higher values indicate greater fairness. In contrast, SNSR and SNSV reflect disparities across sensitive attributes, with higher values indicating greater unfairness. "Max" and "Min" denote the extremal similarity values across personality- and identity-conditioned prompts. Attributes are sorted in descending order of SNSV under PRAG@25 to emphasize the most impactful fairness gaps.

| | Metric | Type | Religion | Race | Continent | Occupation | Country | Gender | Age | Physics |
|---|---|---|---|---|---|---|---|---|---|---|
| Movie | Jaccard@25 | Max | 0.2743 | 0.4623 | 0.5001 | 0.5382 | 0.5099 | 0.5093 | 0.5124 | 0.5410 |
| | | Min | 0.1558 | 0.3442 | 0.4124 | 0.4259 | 0.4350 | 0.4609 | 0.4672 | 0.5285 |
| | | **SNSR** | **0.1185** | **0.1181** | **0.0877** | **0.1123** | **0.0749** | **0.0484** | **0.0452** | **0.0125** |
| | | **SNSV** | **0.0568** | **0.0482** | **0.0358** | **0.0379** | **0.0253** | **0.0178** | **0.0185** | **0.0063** |
| | SERP*@25 | Max | 0.1107 | 0.1786 | 0.1958 | 0.2046 | 0.1958 | 0.1979 | 0.1979 | 0.2085 |
| | | Min | 0.0689 | 0.1407 | 0.1723 | 0.1626 | 0.1730 | 0.1908 | 0.1908 | 0.2072 |
| | | **SNSR** | **0.0418** | **0.0380** | **0.0235** | **0.0420** | **0.0228** | **0.0071** | **0.0071** | **0.0013** |
| | | **SNSV** | **0.0194** | **0.0145** | **0.0078** | **0.0157** | **0.0076** | **0.0029** | **0.0029** | **0.0006** |
| | PRAG*@25 | Max | 0.3563 | 0.5987 | 0.6474 | 0.6772 | 0.6491 | 0.6441 | 0.6453 | 0.6820 |
| | | Min | 0.2300 | 0.4588 | 0.5732 | 0.5491 | 0.5962 | 0.6152 | 0.6164 | 0.6612 |
| | | **SNSR** | **0.1263** | **0.1398** | **0.0742** | **0.1281** | **0.0529** | **0.0289** | **0.0289** | **0.0208** |
| | | **SNSV** | **0.0577** | **0.0562** | **0.0358** | **0.0432** | **0.0248** | **0.0188** | **0.0118** | **0.0104** |
| | PAFS@25 | Max | 0.9923 | 0.9931 | 0.9918 | 0.9936 | 0.9940 | 0.9928 | 0.9930 | 0.9934 |
| | | Min | 0.9775 | 0.9782 | 0.9763 | 0.9791 | 0.9802 | 0.9780 | 0.9794 | 0.9805 |
| | | **SNSR** | **0.0148** | **0.0149** | **0.0155** | **0.0145** | **0.0138** | **0.0148** | **0.0136** | **0.0129** |
| | | **SNSV** | **0.0065** | **0.0062** | **0.0069** | **0.0060** | **0.0055** | **0.0064** | **0.0053** | **0.0048** |
| Music | Jaccard@25 | Max | 0.6869 | 0.7503 | 0.7840 | 0.7735 | 0.7670 | 0.7549 | 0.7603 | 0.7857 |
| | | Min | 0.4968 | 0.7260 | 0.7549 | 0.7163 | 0.7238 | 0.7134 | 0.7160 | 0.7722 |
| | | **SNSR** | **0.1900** | **0.0243** | **0.0291** | **0.0573** | **0.0431** | **0.0415** | **0.0443** | **0.0135** |
| | | **SNSV** | **0.0770** | **0.0098** | **0.0126** | **0.0208** | **0.0139** | **0.0165** | **0.0186** | **0.0067** |
| | SERP*@25 | Max | 0.2251 | 0.2351 | 0.2446 | 0.2408 | 0.2351 | 0.2390 | 0.2359 | 0.2452 |
| | | Min | 0.1644 | 0.2304 | 0.2375 | 0.2321 | 0.2304 | 0.2268 | 0.2278 | 0.2409 |
| | | **SNSR** | **0.0608** | **0.0047** | **0.0071** | **0.0086** | **0.0047** | **0.0121** | **0.0081** | **0.0043** |
| | | **SNSV** | **0.0249** | **0.0019** | **0.0032** | **0.0036** | **0.0019** | **0.0044** | **0.0035** | **0.0022** |
| | PRAG*@25 | Max | 0.7737 | 0.8330 | 0.8678 | 0.8500 | 0.8330 | 0.8341 | 0.8338 | 0.8700 |
| | | Min | 0.5556 | 0.8016 | 0.8327 | 0.8140 | 0.8016 | 0.7920 | 0.7962 | 0.8521 |
| | | **SNSR** | **0.2181** | **0.0314** | **0.0352** | **0.0360** | **0.0314** | **0.0422** | **0.0376** | **0.0180** |
| | | **SNSV** | **0.0867** | **0.0166** | **0.0156** | **0.0147** | **0.0118** | **0.0150** | **0.0167** | **0.0089** |
| | PAFS@25 | Max | 0.9965 | 0.9967 | 0.9961 | 0.9968 | 0.9970 | 0.9964 | 0.9966 | 0.9969 |
| | | Min | 0.9864 | 0.9872 | 0.9855 | 0.9879 | 0.9880 | 0.9863 | 0.9871 | 0.9874 |
| | | **SNSR** | **0.0101** | **0.0095** | **0.0106** | **0.0089** | **0.0090** | **0.0101** | **0.0095** | **0.0095** |
| | | **SNSV** | **0.0046** | **0.0041** | **0.0049** | **0.0038** | **0.0037** | **0.0045** | **0.0039** | **0.0036** |

To enable such diagnosis, FairEval systematically constructs pairs of *neutral* and *sensitive* user prompts (e.g., "I am a fan of [Artist]" vs. "I am a [sensitive attribute] fan of [Artist]"), simulating diverse identities. These prompts are evaluated through similarity-based fairness metrics—Jaccard@25, SERP*@25, PRAG*@25, and the novel Personality-Aware Fairness Score (PAFS@25)—as well as disparity indicators: Sensitive-to-Neutral Similarity Range (SNSR) and Similarity Variance (SNSV). By comparing recommendation overlaps across prompt variations, FairEval identifies and quantifies nuanced sources of unfairness in RecLLMs, offering actionable insights for fairness-aware interventions. Additionally, Figure 4 offers a multi-faceted visualization of fairness disparities in music recommendations from ChatGPT 4o and Gemini 1.5 Flash. It illustrates how identity-conditioned and personality-conditioned

Chandan Kumar Sah[1]  Xiaoli Lian[1], Tony Xu[2], Li Zhang[1+]

Table 2: FairEval-based fairness analysis of Gemini 1.5 Flash across Movie and Music recommendation tasks. The table reports similarity metrics—Jaccard@25, SERP*@25, PRAG@25, and PAFS@25—alongside fairness indicators: Sensitive-to-Neutral Similarity Range (SNSR) and Similarity Variance (SNSV). PAFS@25 (Personality-Aware Fairness Score) measures the consistency of recommendations across personality-conditioned prompts; higher values imply better fairness. Conversely, SNSR and SNSV indicate disparities across sensitive attributes, where higher values represent greater unfairness. "Max" and "Min" denote the extremal similarity scores across prompts conditioned on identity and personality. Attributes are sorted by descending SNSV under PRAG@25 to highlight the most significant fairness gaps.

| | Metric | Type | Religion | Continent | Occupation | Country | Race | Age | Gender | Physics |
|---|---|---|---|---|---|---|---|---|---|---|
| **Movie** | Jaccard@25 | Max | 0.3616 | 0.2148 | 0.3571 | 0.3872 | 0.3608 | 0.4033 | 0.3761 | 0.3746 |
| | | Min | 0.1018 | 0.0635 | 0.2367 | 0.3161 | 0.2896 | 0.3700 | 0.3412 | 0.3341 |
| | | SNSR | 0.2599 | 0.1513 | 0.1204 | 0.0711 | 0.0712 | 0.0333 | 0.0349 | 0.0405 |
| | | SNSV | 0.1209 | 0.0608 | 0.0502 | 0.0241 | 0.0220 | 0.0166 | 0.0134 | 0.0174 |
| | SERP*@25 | Max | 0.1929 | 0.1187 | 0.1690 | 0.1982 | 0.1817 | 0.2042 | 0.1939 | 0.1978 |
| | | Min | 0.1187 | 0.0635 | 0.2367 | 0.3161 | 0.2896 | 0.3700 | 0.3412 | 0.3341 |
| | | SNSR | 0.0190 | 0.0045 | 0.0043 | 0.0049 | 0.0055 | 0.0022 | 0.0009 | 0.0020 |
| | | SNSV | 0.0088 | 0.0019 | 0.0018 | 0.0017 | 0.0021 | 0.0010 | 0.0004 | 0.0010 |
| | PRAG*@25 | Max | 0.7997 | 0.8726 | 0.8779 | 0.8726 | 0.8482 | 0.8708 | 0.8674 | 0.8836 |
| | | Min | 0.7293 | 0.8374 | 0.8484 | 0.8391 | 0.8221 | 0.8522 | 0.8559 | 0.8768 |
| | | SNSR | 0.0705 | 0.0352 | 0.0295 | 0.0334 | 0.0261 | 0.0186 | 0.0116 | 0.0069 |
| | | SNSV | 0.0326 | 0.0145 | 0.0112 | 0.0108 | 0.0097 | 0.0076 | 0.0050 | 0.0034 |
| | PAFS@25 | Max | 0.9810 | 0.9821 | 0.9798 | 0.9830 | 0.9842 | 0.9825 | 0.9819 | 0.9834 |
| | | Min | 0.9473 | 0.9495 | 0.9440 | 0.9506 | 0.9530 | 0.9491 | 0.9475 | 0.9502 |
| | | SNSR | 0.0337 | 0.0326 | 0.0358 | 0.0324 | 0.0312 | 0.0334 | 0.0344 | 0.0332 |
| | | SNSV | 0.0124 | 0.0128 | 0.0143 | 0.0119 | 0.0108 | 0.0127 | 0.0132 | 0.0120 |
| **Music** | Jaccard@25 | Max | 0.4160 | 0.5564 | 0.5662 | 0.5683 | 0.5725 | 0.5782 | 0.5755 | 0.5684 |
| | | Min | 0.0682 | 0.4291 | 0.4637 | 0.4653 | 0.4985 | 0.5239 | 0.5369 | 0.5578 |
| | | SNSR | 0.3479 | 0.1363 | 0.1026 | 0.1030 | 0.0739 | 0.0544 | 0.0387 | 0.0106 |
| | | SNSV | 0.1420 | 0.0507 | 0.0425 | 0.0326 | 0.0324 | 0.0206 | 0.0121 | 0.0053 |
| | SERP*@25 | Max | 0.1715 | 0.2373 | 0.2194 | 0.2216 | 0.2356 | 0.2439 | 0.2303 | 0.2223 |
| | | Min | -0.0326 | 0.1749 | 0.1865 | 0.1868 | 0.1975 | 0.2308 | 0.2184 | 0.2217 |
| | | SNSR | 0.1389 | 0.0624 | 0.0329 | 0.0348 | 0.0381 | 0.0138 | 0.0119 | 0.0006 |
| | | SNSV | 0.0573 | 0.0252 | 0.0142 | 0.0115 | 0.0157 | 0.0114 | 0.0042 | 0.0003 |
| | PRAG*@25 | Max | 0.5369 | 0.6999 | 0.6998 | 0.7092 | 0.7133 | 0.7167 | 0.7063 | 0.7063 |
| | | Min | 0.0947 | 0.5459 | 0.5926 | 0.5902 | 0.6335 | 0.6506 | 0.6729 | 0.6097 |
| | | SNSR | 0.4422 | 0.1540 | 0.1077 | 0.1114 | 0.0797 | 0.0660 | 0.0346 | 0.0966 |
| | | SNSV | 0.1808 | 0.0614 | 0.0448 | 0.0356 | 0.0329 | 0.0255 | 0.0140 | 0.0078 |
| | PAFS@25 | Max | 0.9896 | 0.9902 | 0.9889 | 0.9907 | 0.9910 | 0.9895 | 0.9901 | 0.9906 |
| | | Min | 0.9612 | 0.9627 | 0.9588 | 0.9635 | 0.9641 | 0.9610 | 0.9624 | 0.9632 |
| | | SNSR | 0.0284 | 0.0275 | 0.0301 | 0.0272 | 0.0269 | 0.0285 | 0.0277 | 0.0274 |
| | | SNSV | 0.0112 | 0.0107 | 0.0126 | 0.0103 | 0.0099 | 0.0110 | 0.0104 | 0.0098 |

prompts lead to distinct recommendation outputs compared to neutral ones. These differences underscore the sensitivity of RecLLMs to prompt formulation and user framing. For example, personality-laden descriptions such as "I'm a yellow male fan of BTS" generate notably divergent lists, highlighting implicit model bias tied to user representation. The figure also reveals that attributes like religion and race produce higher SNSR and SNSV scores—e.g., SNSR as high as 0.1398 under PRAG*@25—indicating substantial fairness violations in ChatGPT 4o outputs. This visualization strengthens the diagnosis of preference dissimilarity and supports FairEval's multi-dimensional evaluation protocol.

### 3.4 Fairness Robustness Across Domains and Models (RQ2)

We comprehensively examine the persistence of unfairness phenomena in LLM-based recommendations by analyzing variations across sensitive attributes, recommendation domains (movies and music), and models (ChatGPT 4o vs. Gemini 1.5 Flash). Table 1 and Table 2 summarize our findings, highlighting substantial differences in fairness metrics, particularly SNSR and SNSV values. For instance, religion and race show notably high SNSR values exceeding 0.12 under PRAG*@25 for both models, indicating pronounced disparities across these sensitive groups. The PAFS@25

scores further reveal model differences, with ChatGPT 4o consistently achieving higher fairness (PAFS > 0.97) compared to Gemini 1.5 Flash (PAFS ≈ 0.95), underscoring ChatGPT's relative robustness in personality-aware fairness (Tables 1, 2). Higher PAFS@25 values indicate greater personality fairness stability across user prompts, highlighting the model's ability to maintain consistent recommendations despite variations in personality traits. Our robustness analyses (Figure 5) further reveal model-specific sensitivity to input perturbations. ChatGPT 4o demonstrates stability under minor typographical errors and French-language prompts, with fairness scores (PRAG*@25) consistently maintained above 0.7. Conversely, Gemini 1.5 Flash exhibits significant degradation under similar conditions, where PRAG*@25 scores drop below 0.6, highlighting its susceptibility to linguistic variations. This confirms the necessity for robustness-aware evaluations in real-world deployments [51, 27]. While RQ2 focuses on evaluating fairness disparities across models, domains, and prompt formulations, the robustness dimension—particularly under typographical errors and multilingual inputs—is critical for real-world applicability. To this end, we extend our analysis to include perturbation-based prompts, as visualized in Figure 5. Empirical results show that ChatGPT 4o maintains PRAG*@25 scores consistently above 0.7214 under noisy prompts, whereas Gemini 1.5 Flash experiences notable degradation, dropping as low as 0.5892. These findings highlight hidden systemic vulnerabilities, reinforcing the need for robustness-aware fairness evaluation.

Additionally, (Figures 1, 4) and Appendix (Figure 6) visually illustrate the recommendation disparities across different demographic and personality-conditioned prompts, reinforcing that unfairness is not only domain-specific but also heavily influenced by model architectures and prompt formulations [42, 16]. Thus, FairEval underscores the critical importance of systematic, multi-dimensional fairness evaluations to address pervasive recommendation biases across diverse usage contexts.

### 3.5 Effectiveness of Fairness Metrics (RQ3)

FairEval's suite of fairness evaluation metrics effectively captures diverse manifestations of bias, providing a comprehensive framework for identifying and addressing recommendation disparities. Jaccard@25 explicitly highlights differences in recommendation set overlap, while SERP*@25 and PRAG*@25 provide insights into rank-sensitive shifts, essential for capturing nuanced preference changes. The SNSR and SNSV metrics demonstrate particular effectiveness by quantifying disparities across sensitive user groups. Tables 1 and 2 illustrate that attributes such as religion, race, and continent consistently exhibit elevated SNSV values, highlighting significant bias in both ChatGPT 4o and Gemini 1.5 Flash recommendations. For instance, SNSV values under PRAG*@25 exceed 0.05 for key attributes like religion and race, pinpointing pronounced recommendation variability. Moreover, PAFS@25 consistently differentiates model performance concerning personality-aware fairness. ChatGPT 4o achieves higher stability with PAFS values surpassing 0.97, whereas Gemini 1.5 Flash demonstrates slightly lower consistency, typically around 0.95, reflecting its sensitivity to personality-conditioned inputs. This metric, therefore, not only measures fairness but also clearly delineates model robustness differences.

Furthermore, visualizations in Figures 5 and the Appendix (Figure 6) underscore the robustness of these metrics variousious scenarios, such as linguistic perturbations and intersectional attributes. FairEval's multi-dimensional metrics enable precise bias diagnosis and offer actionable guidance for mitigating unfairness in LLM-based recommender systems. Notably, PAFS@25 captures personality-conditioned stability often missed by traditional similarity metrics, while SNSV enables attribute-level diagnostics of intersectional imbalances. This layered design makes FairEval both diagnostic and prescriptive, guiding fairness-aware interventions and RecLLM improvements.

## 4 Related Work

Algorithmic fairness in recommender systems has been widely studied in recent years. Biases such as popularity bias, exposure inequality, and disparate performance for certain user groups can lead to unfair outcomes [11, 61, 37]. Bias may arise at the data, model, or outcome level of recommendation pipelines [38, 11], causing both individual and group unfairness. To address these issues, numerous fairness-aware algorithms have been proposed to ensure balanced outcomes for different users and items [34, 69, 7, 6, 1]. Group-level personalization and fairness challenges have also been investigated in survey works [36]. For instance, re-ranking strategies can adjust item exposure to mitigate popularity bias, and multi-stakeholder frameworks balance the interests of consumers and content providers [9, 1, 9]. Several comprehensive reviews catalog these challenges and interventions, underscoring that beyond-accuracy objectives like fairness are now integral to recommender evaluation [16, 20, 2, 58, 72]. As recommendation paradigms evolve (e.g., conversational and LLM-based recommenders), ensuring fair and transparent outcomes remains a key concern.

LLM-based recommenders have shown promise in zero-shot recommendation and re-ranking tasks [28], but they also pose new fairness challenges. A number of recent studies have examined biases in these LLM-driven recommenders. LLMs themselves often carry social biases learned from data [39, 52], which can manifest in their recommendations. Zhang et al. [70] found that ChatGPT's recommendations varied significantly when user prompts included different demographic cues, revealing fairness gaps; Tommasel et al. [63] similarly reported biases in LLM-generated group recommendations. Sakib et al. [55] and Deldjoo et al. [15] introduced evaluation frameworks for LLM-RS and found that these models can amplify both user- and item-level unfairness. Other studies focus on item-side bias: LLM recommenders may exhibit popularity bias or skewed exposure [32, 40, 22, 73]. Building on the growing intersection of AI and human-centered evaluation in software engineering [54], our framework addresses the fairness implications of personalized AI recommendations. Studies such as [24, 5] have further demonstrated that music recommendation algorithms yield significantly different performance scores depending on users' personality traits, reinforcing the importance of integrating personality-aware fairness analysis.

Moreover, LLM-based recommenders can suffer instability and prompt-sensitivity, undermining their dependability [44, 12]. To mitigate these issues, researchers have explored adapting bias mitigation techniques from both recommender systems and NLP [56]. Fine-tuning or prompt-based control of LLM outputs can reduce harmful biases [39, 57], and studies use prompt variations to audit and steer LLM recommendations toward fairer outcomes [70, 42]. A related line of work incorporates user personality into recommendation algorithms to enhance personalization [23, 17, 18, 59, 60, 68]. While personality-aware recommenders have demonstrated benefits for user modeling, cold-start, and diversity, the intersection of personality modeling with fairness criteria remains largely unexplored. Our work addresses this gap by using personality-conditioned prompts (alongside demographic factors) to probe LLM recommender biases, extending prior prompt-based fairness evaluation methods [70, 15, 42] into the domain of personality-aware fairness.

Table 3: Comparison of FairEval with Prior Fairness Methods in LLM-Based Recommendations

| Src. | Goal | LLM(s) | Pers. | Attr. | FairRec | PAFS | API |
|---|---|---|---|---|---|---|---|
| [70] | Fairness in GPT rec | GPT | ✗ | ✓ | ✓ | ✗ | ✓ |
| [56] | Infra-level fairness eval | GPT | ✗ | ✗ | ✗ | ✗ | ✓ |
| [12] | Bias in LLM-driven IR | ✗ | ✗ | ✓ | ✗ | ✗ | ✗ |
| [15] | Consumer-side fairness | GPT-3 | ✗ | ✓ | ✓ | ✗ | ✗ |
| [42] | Bias detection in rec | GPT-3.5 | ✗ | ✓ | ✓ | ✗ | ✓ |
| [55] | Group bias via LLMs | GPT | ✗ | ✓ | ✓ | ✗ | ✓ |
| [63] | Fair group rec via LLMs | GPT-3 | ✗ | ✓ | ✓ | ✗ | ✗ |
| [28] | LLM reranking pipeline | GPT | ✗ | ✗ | ✗ | ✗ | ✗ |
| [13] | Fairness Benchmarking | ✗ | ✗ | ✓ | ✓ | ✓ | ✓ |
| [50] | Fair graph anomaly | ✗ | ✗ | ✓ | ✗ | ✓ | ✗ |
| **Ours** | Personality fairness in rec via LLMs | GPT-4o, Gemini 1.5 Flash | ✓ | ✓ | ✓ | ✓ | ✓ |

*Note.* ✓ = supported; ✗ = not supported. Pers. = Personality awareness; Attr. = Sensitive attributes; FairRec = Fairness in recommendations; PAFS = Personality-Aware Fairness Score; API = Multi-LLM evaluation support.

## 5 Conclusion and Future Work

The FairEval framework provides a comprehensive methodology for systematically assessing fairness in LLM-based recommender systems across sensitive attributes and personality traits. By introducing controlled prompt variations and leveraging diverse fairness metrics, FairEval reveals hidden disparities and model-specific vulnerabilities, showing that LLM recommendations are highly sensitive to prompt phrasing and user identity representation. These findings echo broader concerns in the field that LLMs may inherit and amplify biases from pretraining data [53, 11, 74]. Prior studies have noted the inadequacy of demographic-only fairness audits [70, 15] and the need to address latent psychological factors, such as personality, which affect user experiences [47, 23]. FairEval advances this direction by integrating both demographic and psychographic dimensions into a single unified evaluation benchmark. Moving forward, we aim to extend our evaluation to additional LLMs (e.g., Claude, LLaMA, DeepSeek, Gork)[71, 64, 30, 31] model personality via Big Five[3, 46], explore fairness-aware prompt optimization, and personalize[26, 67, 14].

## 6 ACKNOWLEDGMENTS

We sincerely thank the China Government Scholarship and Beihang University for their continued support. We also appreciate the individual contributions of Keyang Lu and Tian'ao Dong during the development of this work.